# Electrical properties of 0.4 cm long single-walled carbon nanotubes

*Shengdong Li[†], Zhen Yu[†], Christopher Rutherglen, Peter J. Burke[*]*

Integrated Nanosystems Research Facility,

Department of Electrical Engineering and Computer Science;

University of California, Irvine, CA 92697

[*] Corresponding author: pburke@uci.edu; [†] These authors contributed equally to this work.

**CORRESPONDING AUTHOR.** Peter J. Burke, Phone number: 949-824-9326, Fax: 949-824-3732. Email: pburke@uci.edu. Mailing address: Electrical Engineering and Computer Science, EG 2232 University of California at Irvine, Irvine, CA 92697-2625

**Abstract.** Centimeter scale aligned carbon nanotube arrays are grown from nanoparticle/metal catalyst pads. We find the nanotubes grow both with and "against the wind". A metal underlayer provides in-situ electrical contact to these long nanotubes with no post growth processing needed. Using the electrically contacted nanotubes, we study electrical transport of 0.4 cm long nanotubes. The source drain I-V curves are quantitatively described by a classical, diffusive model. Our measurements show that the outstanding transport properties of nanotubes can be extended to the cm scale and open the door to large scale integrated nanotube circuits with macroscopic dimensions.

We recently invented a technique to grow mm scale carbon nanotube (CNT) arrays using a methane CVD system[1]. Our technique consisted of a lithographically defined Au underlayer and Fe laden alumina nanoparticles. In this paper, we report four key advances. First, we show that our new growth technique (using a metal underlayer as part of the catalyst) produces *in-situ* electrically contacted nanotubes, not requiring any post-growth processing. Second, by extending the growth time, we extend our technique from the mm to cm scale length CNTs, and grow CNTs up to 0.7 cm in length. Third, by performing electrical transport measurements on several 0.4 cm long CNTs, we show that these nanotubes have outstanding conductivity and mobility over their entire length, as good as the highest mobility micron and sub-mm long nanotubes produced to date. Fourth, we provide new experimental studies to determine whether nanotubes grow upstream or downstream in our CVD system. Interestingly, we find some grow with the wind, but a significant number (of order 50%) of long tubes grow "against the wind". These are the longest individual nanotubes ever contacted electrically by over an order of magnitude. (Durkop et al recently contacted 300 micron long tubes[2].) Our measurement show for the first time that the outstanding electrical properties of nanotubes can be extended to the cm scale, thus opening the door to large scale integrated nanotube circuits with macroscopic dimensions.

The synthesis process is described in detail in reference[1], and is briefly repeated here. CNTs were grown on conducting Si wafers coated with thermal oxide, $t_{ox}$ = 500 nm. The Fe/Mo/alumina nanoparticle catalyst is prepared similar to reference[1]. Alumina nanoparticles (1.0g, Degussa) were dispersed in DI water (250 mL) and then Fe(NO3)3 (1.0g) and MoO2(acac)2 (0.24g) was added to the solution. The mixture was stirred for 24 hours and then sonicated for one hour before applying to Si wafers. Si wafers were first patterned with photoresist. After Cr(50 nm)/Au(200 nm) thin films were deposited on Si wafers using standard thermal evaporation in vacuum, the nanoparticle mixture was then applied. Lift-off in organic solvents gave rise to patterned triple layer catalyst -- nanoparticle/Au/Cr. CNTs were then grown in a home made CVD furnace at 900 C using a mixture of CH4 (1000 sccm) and H2 (200 sccm). Our synthesis technique is distinct from that developed by Liu[3-6], in that our technique does not require fast heating or a vertical electric field, and is thus simpler. An additional difference is that our technique provides in situ electrical contact, as we show below.

It is a non-trivial challenge to characterize nanostructures with high aspect ratios such as our CNTs. We developed a technique to image the long nanotubes using an SEM (Hitachi 4700). SEM images were taken at low acceleration voltage (1 kV). High acceleration voltages were not used because SEM imaging of CNTs on an insulating substrate is based on conductance contrast or potential contrast[7]. In our work, in order to enhance the contrast, we first charge the catalyst pad that connects the CNTs by imaging only the catalyst pad. Within less than 30 seconds, the catalyst pad, which is a conductive metal, becomes negatively charged, as are the CNTs which are in electrical contact with the pads. Next, the SEM is switched to low magnification mode. CNTs connected to catalyst pads were imaged because their potential is different from the oxide underneath. We specifically designed the catalyst pad geometry to fit the field of view of the Hitachi 4700 SEM at low magnification. The distance between neighboring catalyst pads is 4 mm. This technique dramatically enhances the contrast of the SEM images and allows fast imaging of these centimeter scale nantoubes.

In our former work[1], mm long CNTs were grown 15 minutes. In this work, we extend that growth time to 70 minutes. Based on this long growth time we are able to synthesize CNTs up to 0.7 cm in length. In figures 1 and 2, we show SEMs image of sample "A" and "B", both individual CNTs bridging two catalyst/electrode pads, of length 0.4 cm. High magnification images (not shown) indicate only one nanotube under SEM imaging, although some short nanotubes perpendicular to the long tube are occasionally present. It is interesting to see that, in Figure 1, CNTs grow both in the direction of gas flow, and also against the direction of gas flow, i.e. "against the wind". In figure 3, we show an isolated catalyst/electrode pad, which clearly has nanotubes growing both with and against the wind. We studied the growth direction of CNTs from 24 catalyst pads, and found growth both with and against the wind of CNTs > 0.5 mm in length for all of the pads. Careful study of similar catalyst patterns under similar growth conditions but shorter active growth times (usually 10-15 minutes) also produce mm scale CNTs both with and against the wind. Thus, figures 1 and 3 are not isolated events, but represent a systematic trend observed in our experiments. Additionally, while the nanotubes growing against the wind in figures 1 and 3 seem to be straighter than those growing with the wind, other images (not shown) indicate that this is not a systematic trend: We have not observed any clear systematic difference between the "straightness" or any other quality of nanotubes that grow with or against the wind. While an explanation of this result is currently lacking, it suggests that the growth mechanism is not simply a catalyst nanoparticle "blowing in the wind", with the nanotube growing behind it. Future proposed CNT synthesis mechanisms will need to explain and be severely constrained by this simple experimental observation. Our studies indicate that growth terminates either due to the presence of an obstruction (such as a neighboring catalyst/electrode pad), the edge of the wafer, or due to stopping the flow of methane. Therefore, even for these cm scale CNTs, we have not yet reached the fundamental limit for the length, and indeed we have no evidence that there is any limit at all.

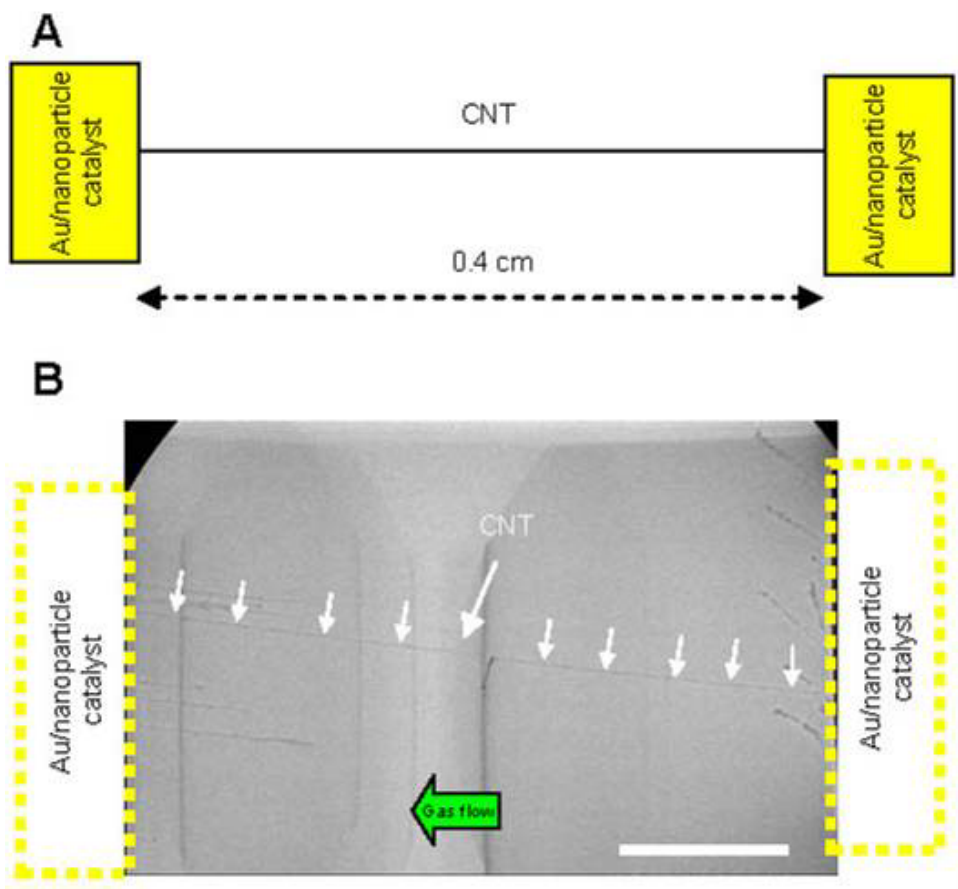

**Figure 1:** A) Schematic image of catalyst and CNT geometry. B) SEM image of individual CNT (sample A) bridging a 0.4 cm gap between two catalyst/electrode pads. Scale bar: 1 mm.

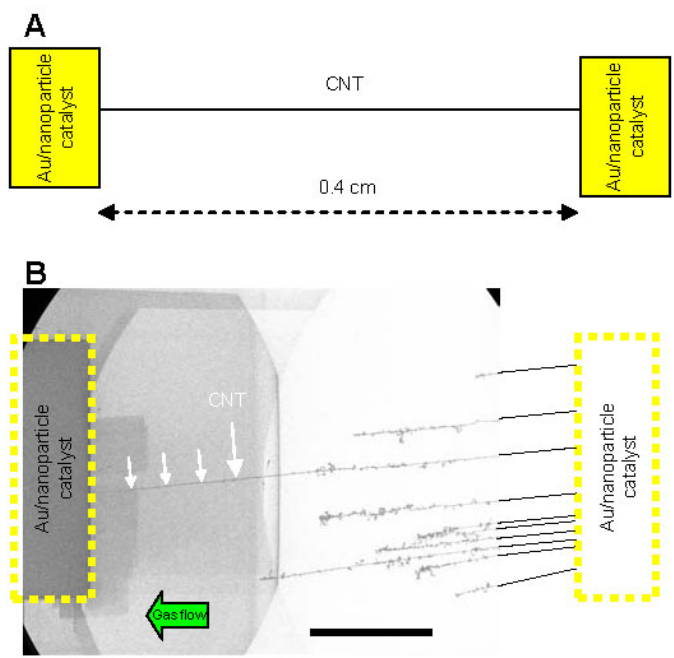

**Figure 2 A) Schematic image of catalyst and CNT geometry. B) SEM image of individual CNT (sample B) bridging a 0.4 cm gap between two catalyst/electrode pads. Scale bar: 1 mm.**

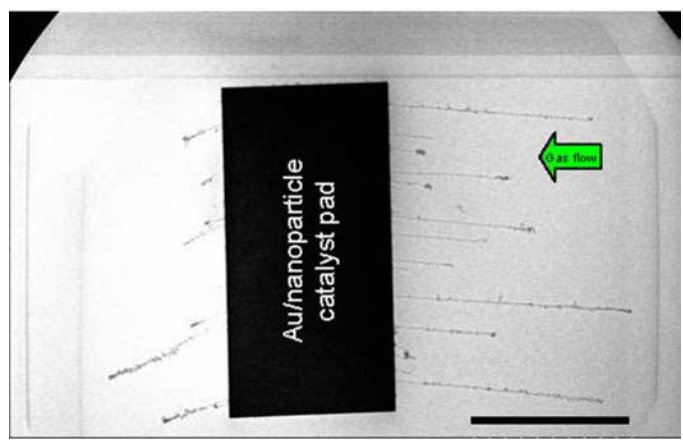

**Figure 3: SEM image showing nanotube growth both with and "against the wind". Scale bar: 1 mm.**

Since the catalyst pattern consists of a metallic (Au) underlayer, it can also be considered as an electrode, hence our nomenclature: catalyst/electrode pad. In order to test this hypothesis, we used a simple probe station with sharp metal tips to contact the left and right catalyst/electrode pads shown in Figure 1. Our results indicate that the catalyst pads can indeed be used as electrodes with no post-growth processing needed, hence the term *in-situ* electrical contact.

We systematically studied a total of 66 neighboring electrode/catalyst pairs with 0.4 cm gaps of the same geometry on the two different wafers. Of these pairs, 63 had no nanotube bridging the gap visible under SEM, 2 pairs had one nanotube bridging the gap under SEM, and 1 pair had 2 nanotubes bridging the gap under SEM (see Supplemental Information). In all cases where no nanotube was bridging the gap, the small bias resistance was found to be greater than 3 GΩ. The two pairs with one nanotube bridging the gap are presented as Sample A and B below, and clearly show conductance of order 20 nS between electrode pads. The pair with two nanotubes bridging the gap exhibited a depletion curve (using the substrate as a gate) with two different turn-on voltages, indicating that both nanotubes are electrically contacted and semiconducting, each with a different threshold voltage. Additional AFM studies (see Supplemental Information) indicate the diameter of our CNTs is less than 5 nm, and that the substrate is atomically smooth (i.e. free of amorphous carbon) even after growth. Taken together, these systematic studies indicate that our samples A and B are indeed electrically contacted, individual CNTs.

We now turn to more detailed electrical studies of our electrically contacted nanotubes. In figure 4, we show the depletion curve (low bias conductance vs. gate voltage) measured at room temperature in ambient air for sample A when the substrate is used as a gate. These measurements were performed on as-grown CNTs, with no post-growth processing. We find the carbon nanotube to be p-type, similar to previous studies on much shorter nanotubes[8]. Hysteresis is also evident in the depletion curve, presumably due to trapped charges in the oxide; similar hysteresis is also observed in our shorter nanotubes, and so is not unique to our cm scale CNTs. The off-state conductance (0.5 nS) indicates an on-off ratio of 40. The conductance in the off state is currently not understood. It is possible that our nanotube is double walled, and that one of the shells is metallic, albeit highly resistive. Our second sample (sample B) exhibited a similar depletion curve, with an on conductance of 35 nS, an off conductance of 0.1 nS, and an on/off ratio of 300.

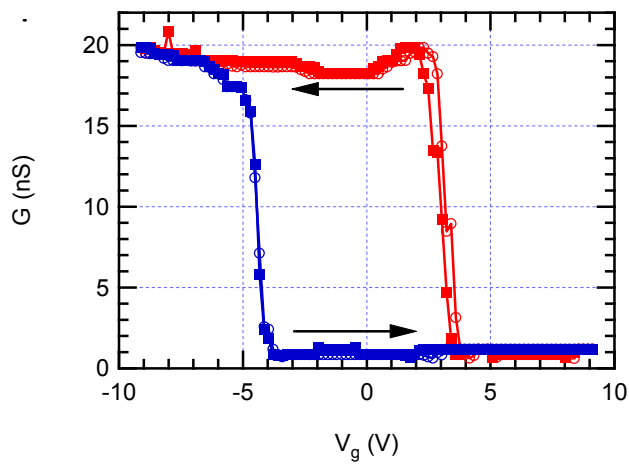

**Figure 4 Depletion curve for sample A. (30 minute each scan direction.) The squares and circles correspond to separate sweeps, and show that the depletion curve exhibits reproducible hysteresis.**

The measured resistance for our devices corresponds to a two-terminal measurement. A key question which naturally arises in this context is whether the measured resistance is dominantly due to the contacts, or to the intrinsic resistance of the nanotube itself. In order to perform additional characterization of the resistances or Schottky barriers at the contacts, we have performed a series of experiments on CNTs synthesized with our technique of length 20 microns, much shorter than the nanotubes studied above. If the contact resistance dominates our long tubes, then the resistance for short and long tubes should be the same. On the other hand, if the contact resistance for the 0.4 cm long tubes is negligible compared to the bulk resistance, than the resistance of much shorter 20 micron tubes should be much lower. Thus, by measuring the conductance of these shorter tubes, we are able to better determine the contributions of the contact resistance and bulk resistance.

In these new experiments, large catalyst pads (1 mm x 1 mm) separated by a gap of 20 microns were used to grow CNTs. The depletion curve of the nanotubes bridging the gaps was measured after growth, with no post-growth processing. Catalyst pad pairs with only one nanotube bridging the gap (as determined by SEM imaging) were studied. The depletion curves for 3 different catalyst pad pairs

8/13/2004

measured at room temperature in air exhibited p-type behavior, similar to the 0.4 cm CNTs. In our experiments, whenever the nanotubes bridge the gap as visible under SEM, they are in electrical contact with the catalyst pads. *This indicates that the in-situ contact is reliable, robust, and reproducible.* The contact arrangement may at first glance seem "unusual." The mechanism of the electrical contact is most likely simple melting of the gold underlayer during CVD growth, which flows over the nanotube just after growth during the CVD run. If this picture is correct, then the contact arrangement is not unusual at all, but similar to other contact arrangements where the nanotube is contacted after growth via evaporated gold, then annealed.

The contact resistance is expected to vary from nanotube-to-nanotube. For our 20 micron long tubes, the lowest resistance measured is 150 kOhms, and the highest resistance measured is 6 Mohm. In control experiments performed in our lab on post-growth contacted CNTs using evaporated Ti/Au electrodes[9], we found CNTs as low as 80 kOhm of length 10 microns. Therefore, the contact resistance of our in-situ contacted nanotubes can be as low as 15 kOhm. This indicates that the 0.4 cm long CNTs (R ~ 50 Mohm) are dominated by the bulk resistance, and that the contact resistance is insignificant for those tubes.

An interesting question, which we are now in a unique position to answer, is what is the resistance vs. length for a carbon nanotube, in a case where the contact resistance is negligible? For very short CNTs which are in the ballistic limit, the contact resistance dominates. For our long CNTs, the contact resistance is negligible. We plot in figure 5 the resistance vs. length for our long and short nanotubes, as well as some key results from the literature[2, 9, 10]. Our results are consistent with a resistance per unit length of 6 kOhm/micron. This scaling behavior is consistent over 4 orders of magnitude in both length and resistance.

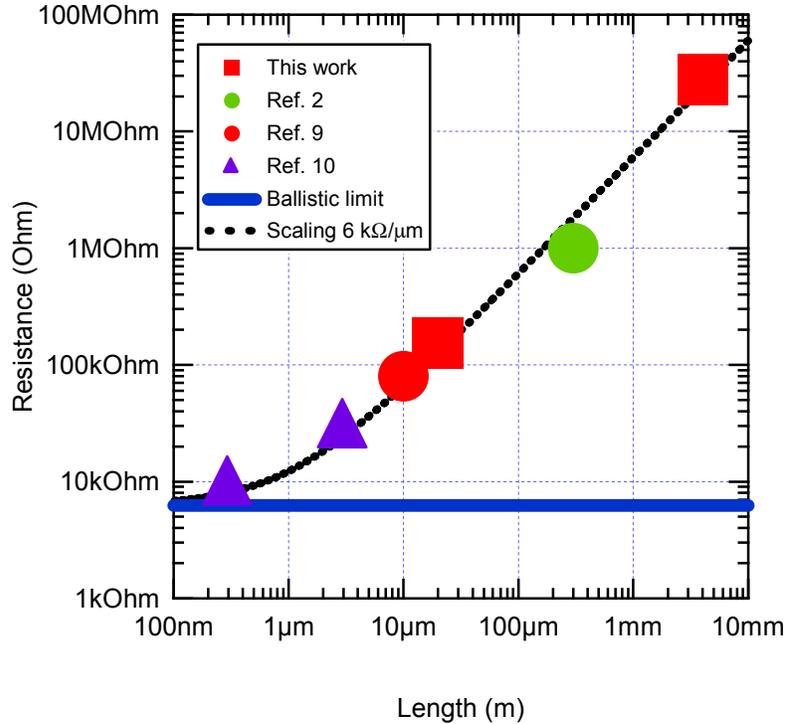

**Figure 5: Resistance vs. length for this work and related references. All data are at room temperature.**

For our long tubes, the on state resistance is dominantly due to the channel resistance, and not the contact resistance. For sample B this would imply a resistance per unit length of ~ 7 kΩ/μm, a 1d conductivity of 1.4 x $10^{-8}$ Ω-cm, and a mean-free-path (= conductivity/ 2 $G_0$, with $G_0$ the conductance quantum) of 1 μm, comparable to the best measured nanotube conductivities for both metallic[8] and semiconducting[2] carbon nanotubes to date. Even a short, defective region of our cm scale CNTs would cause the measured resistance to be dramatically higher. This leads us to the extraordinary conclusion that our cm scale CNTs are as defect free over the entire length of the tube as the best CNTs synthesized to date.

We next turn to the high bias I-V curves of the nanotubes. In figure 6, we plot I-V curves for sample B for a series of gate (i.e. substrate) voltages. Because of the length of our nanotubes, the electric field is much less than the predicted value at which velocity saturation occurs. In our work the electric field strength is ~ $10^2$ V/m; velocity saturation is predicted[11] at ~ $10^6$ V/m. In order to analyze the result quantitatively, we have performed two-parameter fits of the I-V curves to the following 1d analog of a MOSFET I-V curve in the case where the electric field is always low enough to consider the mobility a constant, independent of the electric field:

$$(1) \quad I_{D-S} = \frac{\mu C}{L}\left[(V_G - V_T)V_D - \frac{V_D^2}{2}\right]$$

Here, μ is the mobility, C the capacitance per unit length, L the length, $V_G$ the gate voltage, $V_D$ the drain voltage, and $V_T$ the threshold voltage. This curve is expected to hold below the saturation region. Our resultant two-parameter curve fits (shown as dotted black lines in Figure 6) are in excellent quantitative agreement with the measured I-V curves. Using the results of the curve fit, we can extrapolate a value for the mobility and threshold voltage, assuming we know the capacitance per unit length. For estimation purposes, the capacitance per unit length can be taken to be 200 aF/μm[2, 12]. Using this estimate, and the $V^2$ coefficient from the fitted I-V curves, we can determine a mobility estimate of 5000-20000 cm$^2$/V-s. This is comparable to the highest mobility room temperature semiconductors available (e.g. GaAs, InP), and similar to previous work on nanotubes an order of magnitude shorter[2].

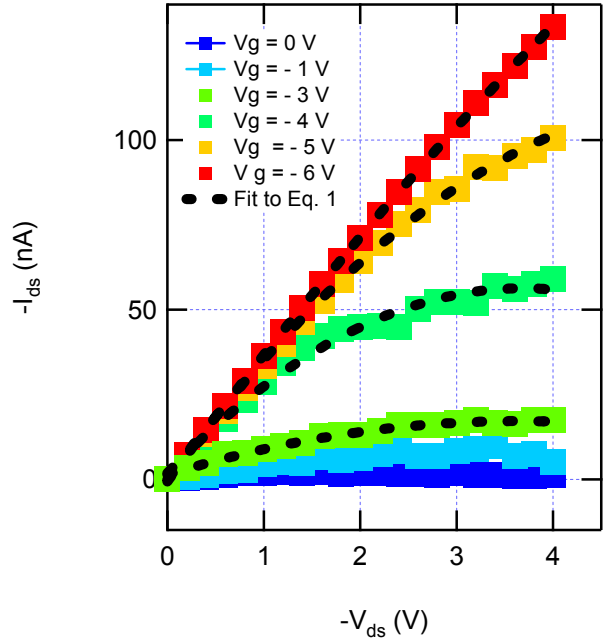

**Figure 6 Source-drain I-V curve for sample B.**

Recently the issue of the mechanism of transistor action in nanotube transistors has been discussed by several groups. Because we have very long nanotubes, our data may provide some useful insight into this discussion. Heinze[13] has presented arguments that the sole effect of a gate is to modulate the Schottky-barrier contact resistance, not the bulk resistance, implying that a carbon nanotube transistor is a Schottky barrier device. However, these arguments generally considered short (micron length) carbon nanotubes. Javey[10] has recently developed contact technology to significantly suppress the Schottky barrier contact resistance. However, attempts to fit their measured I-V curves to equation 1 failed, presumably because Javey's experiments on short (3 micron long) nanotubes were in the high-electric field regime, where equation 1 is no longer applicable. Javey's electric fields were of order $10^6$ V/m. Durkop[2] recently studied 300 micron long nanotubes, and concludes that they are classical field effect transistors. Since our nanotube transistor I-V curves apparently quantitatively satisfy equation 1 at room temperature, we are lead to the tentative conclusion that our work is one demonstration that the mechanism for nanotube transistor action, for cm scale nanotubes, is classical field effect. How this relates to the shorter nanotube transistors with either large or small (ballistic) contact resistances will be an important issue for future study, which will be guided by our data in the limit of extremely long tubes.

In conclusion, we have synthesized and measured the electrical properties of 0.4 cm CNTs. The significance of these studies is multifold. First, it is a fundamental science study showing for the first time that nanometer diameter wires can conduct electricity on centimeter length scales, which heretofore is unprecedented. These results allow other researchers to consider applications of nanotubes at the cm scale for the first time. In the area of electronics, cm scale nanotubes arrays could be used as ultra-dense memory storage. In the area of materials, our studies could have impact on the electrical properties of high-strength composite materials incorporating cm long nanotubes; our results that the electrical defect density is low may indicate that the mechanical defect density is low as well, even for cm long tubes. Just as we have not yet reached the limit for the length of carbon nanotube growth using our synthesis technique, we also have not yet reached the limit of their sphere of application.

**Acknowledgement.** This work was supported by the Army Research Office (award DAAD19-02-1-0387), the Office of the Naval Research (award N00014-02-1-0456), and DARPA (award N66001-03-1-8914), and the National Science Foundation (award ECS-0300557). We thank Jake Hess of the INRF for assistance with construction and design of the CVD system.

Supplemental Information:

1) AFM images.

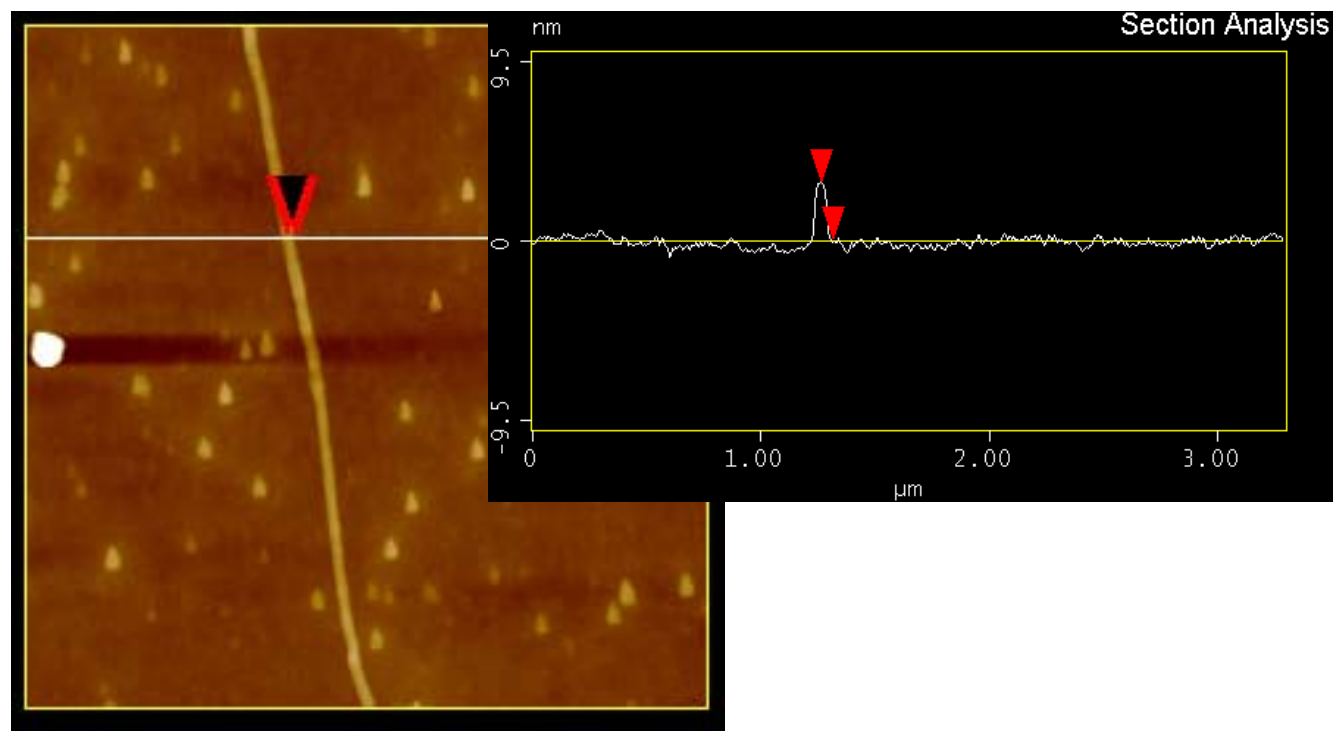

Figure S1 AFM height image of a CNT (length > 1mm), diameter = 3 nm,

Figure S1 is an AFM image of an as grown CNT from the same growth run that produced the 0.7 cm CNT. The CNT is longer than 1 mm and the diameter is 3 nm. The large white dot on the left edge is a dust attached during imaging. The other smaller particles are inactive nanoparticles. Zoom-in images (not shown) reveal that the substrate to be atomically smooth, which indicates that our growth process is free of amorphous carbon or at least negligible, even though the growth time is > 60 minutes.

2) SEM image of 2 nanotubes bridging gap

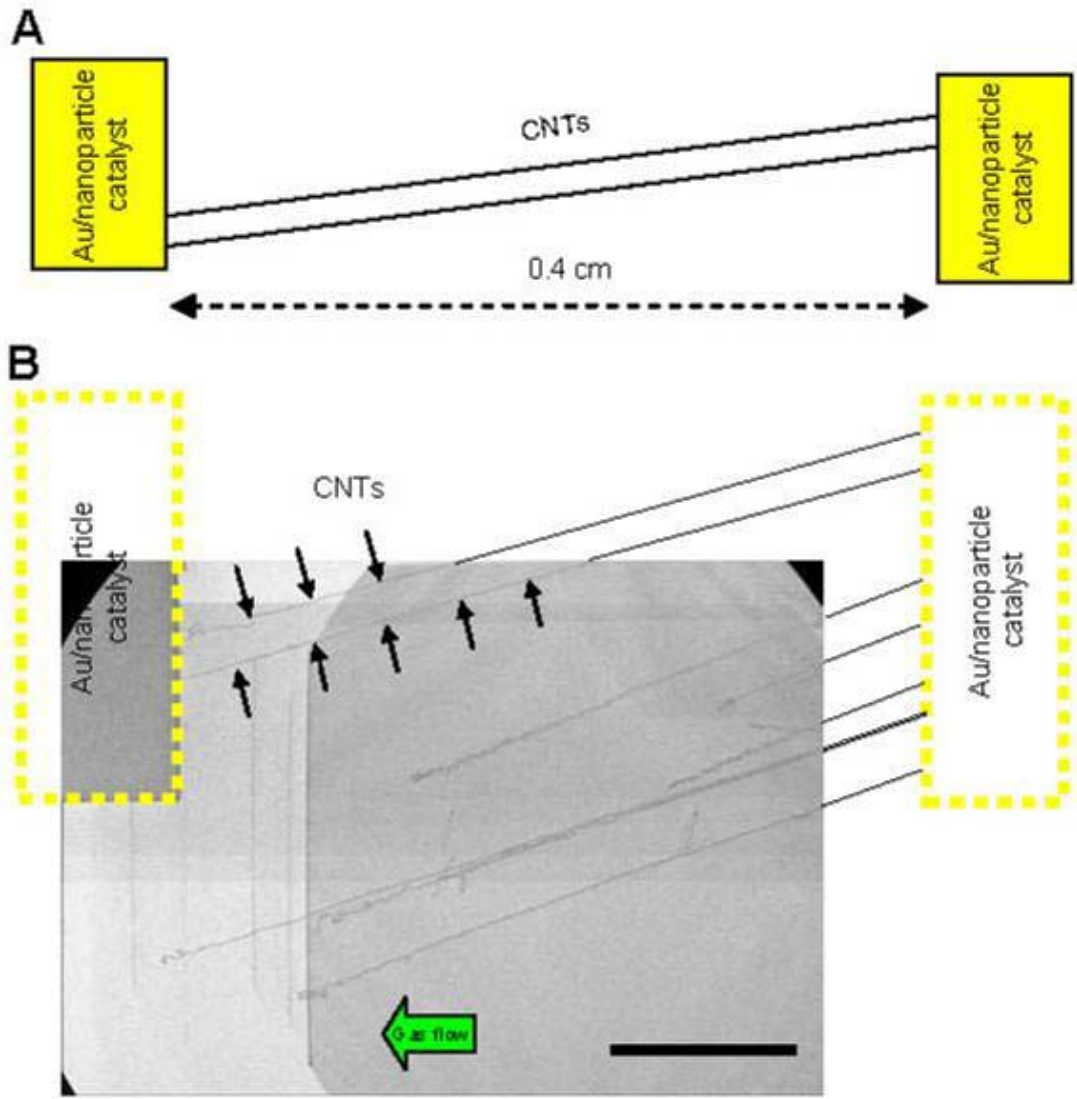

**Figure S2: SEM image of two nanotubes bridging gap.**

The depletion curve for this sample showed two turn-on voltages, indicating that each nanotube is electrically contacted, is semiconducting, and has a separate threshold voltage.

3) SEM image of long nanotube growing "against the wind"

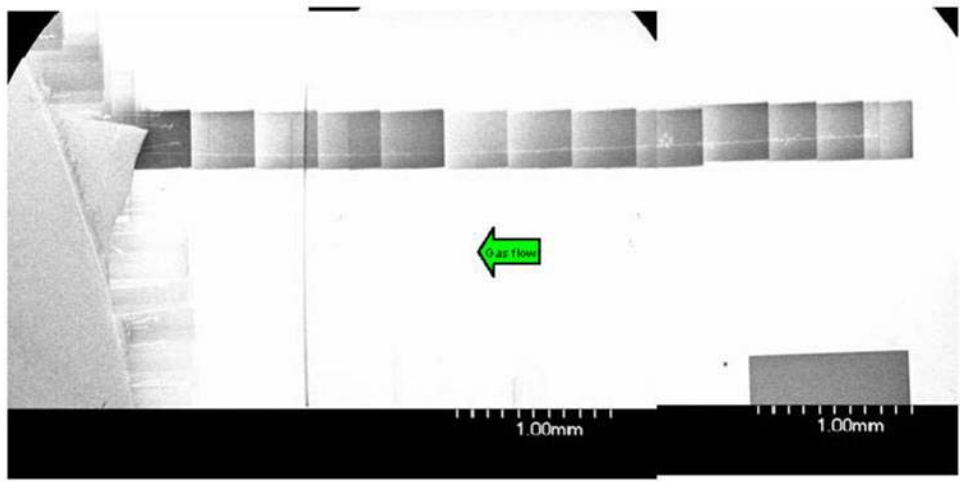

**Figure S3: SEM image of 0.5 cm long nanotube. The dark regions are charged from previous SEM imaging at high magnification.**